
\input phyzzx
\baselineskip=18pt
\hfill {IP-ASTP-07}\break
\strut\hfill {CU-TP-701}\break
\FRONTPAGE
\strut\hfill {SNUTP-95-067} \break 
\font\twelvebf=cmbx12 scaled\magstep2

\vskip 0.0in
\centerline {\twelvebf The Chern-Simons Coefficient in }
\centerline {\twelvebf Supersymmetric Yang-Mills Chern-Simons Theories}

\vskip .5in
\centerline{{\it Hsien-Chung Kao} }
\vskip .1in
\centerline{ Institute of Physics, Academia Sinica}
\centerline{ Nankang, Taipei, 11529 Taiwan}
\vskip 0.3in
\centerline{{\it  Kimyeong Lee }}
\vskip .1in
\centerline{ Physics Department, Columbia University}
\centerline{ New York, N.Y.  10027, U.S.A.}
\vskip 0.3in
\centerline{ {\rm and}}
\vskip .1in
\centerline{ \it Taejin Lee}
\vskip .1in
\centerline{ Physics Department, Kangwon National University}
\centerline{ Chuncheon, 200-701, Korea}
\vskip 0.4in
\vskip 0.1in

We study one-loop correction to the Chern-Simons coefficient
$\kappa=k/4\pi$ in $N=1,2,3$ supersymmetric Yang-Mills Chern-Simons
systems.  In the pure bosonic case, the shift of the parameter $k$ is
known to be $k\rightarrow k + c_v$, where $c_v$ is the quadratic
Casimir of the gauge group. In the $N=1$ case, the fermionic
contribution cancels the bosonic contribution by half and the shift is
$k \rightarrow k+ c_v/2$, making the theory anomalous if $c_v$ is odd.
  In the $N=2,3$ cases, the fermionic contribution cancels the bosonic
contribution completely and there is no correction.  We also calculate
the mass corrections, showing the supersymmetry is preserved.  As the
matter fields decouple from the gauge field in the pure Chern-Simons
limit, this work sheds some light on the regularization dependency of
the correction in pure Chern-Simons systems. We also discuss the
implication to the case when the gauge symmetry is spontaneously
broken by the Higgs mechanism.

\vfill \endpage


\def\pr#1#2#3{Phys.  Rev.  {\bf D#1}, #2 (19#3)}
\def\prl#1#2#3{Phys. Rev.  Lett.  {\bf #1}, #2 (19#3)}

\def\np#1#2#3{Nucl.  Phys.  {\bf B#1}, #2 (19#3)}
\def\pl#1#2#3{Phys.  Lett.  {\bf B#1}, #2 (19#3)}
\def\ibid#1#2#3{ {\it ibid.} {\bf #1}, #2 (19#3)}

\def\tr{\,{\rm tr}}

\def\dsl{{{\rm D} \hskip -0.65em {\it /}}}
\def\fsl{{{\rm B} \hskip -0.65em {\it /}}}
\def\psl{{{\partial} \hskip -0.6em {\it /}}}



\REF\rQC{S.  Deser, R.  Jackiw and S.  Templeton, \prl{48}{975}{82};
S.  Deser, R.  Jackiw and S.  Templeton, Ann.  Phys. (N.Y.)  {\bf 140},
372 (1982).} 

\REF\rPisarski{R.D.  Pisarski and S.  Rao, \pr{32}{2081}{85}.}

\REF\rALL{W. Chen, G.W. Semenoff and Y.-S. Wu, Mod. Phys. Lett.
{\bf A 5}, 1833 (1990); L. \'Alvarez-Gaum\'e, J.M.F. Labastida and A.V.
Ramallo,  \np{334}{103}{90}; M. Asorey and F. Falceto, \pl{241}{31}{90};
C.P. Martin, \pl{241}{513}{90}; M.A. Shifman, \np{352}{87}{91}.}

\REF\rNoco{E. Guadagnini, M. Martellini and M. Mintchev,
\pl{227}{11}{89}.} 


\REF\rAsorey{M. Asorey, F. Falceto, J.L. L\'opez and G. Luz\'on,
\np{429}{344}{94}.} 

\REF\rKaob{H.-C. Kao and K. Lee, \pr{46}{4691}{92}; H.-C. Kao,
\ibid{50}{2881}{94}.} 

\REF\rIvanov{S.J.  Gates, M.T. Grisaru, M. Rocek
and W. Siegel, ``Superspace or One Thousand and One
Lessons in Supersymmetry,'' Benjamin/Cummings, Reading (1983);
E.A. Ivanov, \pl{268}{203}{91}; S.J. Gates and H. Nishino,
\ibid{281}{72}{92}.} 

\REF\rWitt{E. Witten, \pl{117}{324}{82}.} 

\REF\rWitten{E. Witten, Comm. Math. Phys. {\bf 117}, 353 (1988).}

\REF\rKhle{S. Yu. Khlebnikov and M. Shaposhinikov,
\pl{254}{148}{91}.}

\REF\rChen{L. Chen, G. Dunne, K. Haller and E. Lim-Lombridas,
\pl{348}{468}{95}.}

\REF\rKharea{ A. Khare, R. MacKenzie and M. Paranjape,
\pl{343}{239}{95}.}%

\REF\rKaoa{H-C. Kao, K. Lee, C. Lee and T. Lee, \pl{341}{181}{94}.}

There has been recently some interest in the regularization dependency
of the quantum correction to the Chern-Simons coefficient. For the
quantum theory to be invariant under the large gauge transformations,
the Chern-Simons coefficient $k$ should be an integer[1]. In the
Yang-Mills Chern-Simons theories, it is known that due to quantum
correction the coefficient is
shifted from $k  \rightarrow k + c_v$[2,3]. If we take a limit
where the Yang-Mills term disappear, we could conclude that there is a
nonzero correction even in the pure Chern-Simons theory. However, the
naive calculation in the pure Chern-Simons theory does not show such a
correction[4]. Also we can use a somewhat complicated regularization
to get many different corrections to the coefficient[5]. In short, there
seems to be no clear principle to determine the quantum correction to
the Chern-Simons coefficient in the pure Chern-Simons theory.

In this paper we study the $N=1,2,3$ supersymmetric Yang-Mills Chern-Simons
systems.  It is known that the maximal supersymmetry is $N=3$ as the
massive vector multiplet can carry spin $1,1/2,0,-1/2$ up to the sign[6].
As the Chern-Simons coefficient also appears in the mass term for the
matter fields, the supersymmetry puts an additional constraint on the
quantum correction to the parameter $\kappa$.  Since one-loop corrections
in the supersymmetric theories turn out to be finite without any
regularization, there is no regularization dependency in the quantum
correction.  Interestingly, the matter part of the supermultiplet decouples
the gauge field in the pure Chern-Simons limit.  Thus, the correction in
the pure Chern-Simons theory depends on whether there are other matter
fields even though there is no interaction between them.  Of course this is
another disguise of the regularization dependency.  However, we have an
additional structure, supersymmetry, in the theory which we should respect,
and so obtain a unique correction for each given supersymmetry.  We should
emphasize the correction to the Chern-Simons coefficient is meaningful only
if there is any physical process which explores the physics of distance
shorter than the Compton wavelength of the gauge bosons.

One-loop corrections, especially due to tadpole diagrams, can be linearly
divergent and disappear in the dimensional regularization.  In our
supersymmetric models the linear divergences explicitly cancel each other,
making one-loop corrections finite even before the regularization.  While the
superfield formalism is more efficient, here we choose to calculate the
one-loop
corrections in the components to compare it with the known result by Pisarski
and Rao for the bosonic part [2].  Their work was done in Euclidean time with
the dimensional regularization and our work is done in Minkowski time.  Our
results agree with theirs where they overlap.

The plan of this paper is as follows. We introduce the $N=3$
supersymmetric Lagrangian and its supersymmetric transformations.
Straightforward truncations lead to the $N=1,2$ models. Then we study
one-loop corrections in general and establish the method to calculate  the
renormalized coupling constants and masses. After that we perform  the
relevant one-loop calculations and present the renormalized
parameters.  Finally we discuss the implications of our results to the
pure Chern-Simons theories and to the quantum correction in the broken
phases of the  self-dual Chern-Simons Higgs theories.

First, we present the $N=3$ supersymmetric Yang-Mills Chern-Simons theory
of a given gauge group with an explicit $O(3)$ symmetry.  The gauge
multiplet is made of the field $A_\mu$ for one massive vector of spin $1$,
and the field $\lambda_a$ for three Majorana fermions of spin $1/2$, and
the field $C_a$ for three neutral scalar bosons and the field $\chi$ for a
Majorana fermion of spin $-1/2$.  The super Yang-Mills Lagrangian can be
obtained from the dimensional reduction of the pure $N=2$ super Yang-Mills
theory in four dimensions,
$$ \eqalign{ {\cal L}_{YM} = {1\over g^2}\tr \biggl\{ & -{1
\over 2}F^2_{\mu \nu} + (D_\mu C_a)^2 + (D_a)^2 +
i\bar{\lambda}_a\dsl_\mu\lambda_a + i\bar{\chi}\dsl_\mu\chi \cr & +
i\epsilon_{abc}\bar{\lambda}_a[ \lambda_b,C_c] - 2i\bar{\lambda}_a[\chi,
C_a] - {1 \over 2}[C_a, C_b][C_b, C_a] \biggr\}, \cr }
\eqno\eq $$
where
$D_\mu = \partial_\mu - i[A_\mu, ]$, and $a,b,c = 1,2,3$.  The gauge group
generators in the fundamental representation satisfy $[T^m, T^n] =
if^{lmn}T^l,$ and $\tr T^mT^n = \delta^{mn}/2$.
The fields belong to the adjoint representation and $A_\mu = A_\mu^m T^a$ et.
cetra.  The quadratic Casimir
number $c_v$ for the group is given by $f^{kmn}f^{lmn} = c_v \delta^{kl}$.
Here, the metric is given by $\eta_{\mu\nu} = {\rm diag}(1,-1,-1)$ and
$\epsilon^{012} = \epsilon_{012} = 1$.  The pure imaginary gamma matrices
satisfy $\gamma^\mu \gamma^\nu = \eta^{\mu\nu} -i\epsilon^{\mu\nu\rho}
\gamma_\rho$.  The dimensional reduction of the $N=2$ supersymmetry implies
that this theory has the not-so-obvious $N=4$ supersymmetry.

By supersymmetrizing the Chern-Simons term, we get the  supersymmetric
Chern-Simons Lagrangian[6],
$$ {\cal L}_{CS} =   \kappa \; \tr \biggl \{ \epsilon^{\mu\nu\rho}
(A_\mu \partial_\nu A_\rho - {2\over 3}iA_\mu A_\nu A_\rho)
- \bar{\lambda}_a\lambda_a + \bar{\chi}\chi
 +  2C_a D_a + {i\over 3}\epsilon_{abc}C_a[C_b,C_c]\biggr\}. \eqn\LYMCS
$$
The Lagrangian we study is the sum of ${\cal L}_{YM} $ and ${\cal
L}_{CS}$[6].  Note that the corresponding $N=2$ theories can be easily
obtained by setting $C_1 = C_2 = D_1 = D_2 = \lambda_3 =
\chi = 0$. The $N=1$ theories are obtained from $N=2$ theories by further
imposing $C_3=0, \lambda_2=0$. The $N=3$ supersymmetric transformation
of the fields is
given by
$$ \eqalign{ \delta A_\mu &\ =-i \bar{\alpha}_a \gamma_\mu \lambda_a,
\cr
\delta \lambda_a &\ = i \fsl \alpha_a - \epsilon_{abc} (D_b \alpha_c
-i \dsl\, C_b \alpha_c ) + i[C_a,C_b]\alpha_b , \cr
\delta \chi &\ = -i\dsl\, C_a \alpha_a - D_a \alpha_a + {i\over 2}
\epsilon_{abc}
[C_b,C_c] \alpha_a , \cr
\delta C_a &\ = - \epsilon_{abc} \bar{\alpha_b}\lambda_c +
\bar{\alpha}_a\chi , \cr
\delta D_a &\ = i \epsilon_{abc} \bar{\alpha}_b \dsl\, \lambda_c + i
\bar{\alpha}_a \dsl\, \chi +i[\bar{\alpha}_b \lambda_a, C_b] \cr
&\ \;\;\;\;\;\;
-i[\bar{\alpha}_b \lambda_b, C_a] +i [\bar{\alpha}_a \lambda_b, C_b]
- i\epsilon_{abc}\bar{\alpha}_b[\chi,C_c], \cr}
\eqn\sutr $$
where $B^\mu = \epsilon^{\mu\nu\rho}\partial_\nu A_\rho$.  While we do
not know whether there is a $N=3$ superfield
formalism, clearly it can be written in terms of the $N=1$ or $N=2$
superfields[7].

By using the field equation $D_a + \kappa g^2C_a = 0,$ of the
Lagrangian ${\cal L}_{YM} + {\cal L}_{CS}$, we can eliminate the
auxiliary field $D_a$ and obtain the following on-shell Lagrangian:
$$
\eqalign{ {\cal L} = {1\over g^2}\tr \biggl\{
& -{1 \over 2}F^2_{\mu \nu} + (D_\mu C_a)^2
+ i\bar{\lambda}_a\gamma^\mu D_\mu\lambda_a
+ i\bar{\chi}\gamma^\mu D_\mu\chi \cr
& + i\epsilon_{abc}\bar{\lambda}_a[ \lambda_b,C_c]
- 2i\bar{\lambda}_a[\chi, C_a]
- {1 \over 2}[C_a, C_b][C_b, C_a] \biggr\} \cr
+ \kappa \; tr \biggl \{& \epsilon^{\mu\nu\rho}
(A_\mu \partial_\nu A_\rho - {2\over 3}iA_\mu A_\nu A_\rho) - \kappa g^2 C_a^2
- \bar{\lambda}_a\lambda_a + \bar{\chi}\chi \cr
& - {i\over 3}\epsilon_{abc}C_a[C_b,C_c]\biggr\}. \cr}\eqn\OnShL
$$
Due to the mass term for $C_a$, the ground state of the
theory is  the symmetric vacuum where  $<C_a>_{v} = 0 $.
 This contrasts the pure Yang-Mills case whose vacua have a flat direction.

We are here interested in calculating one-loop corrections to the theory. If
we scale the gauge field by $A^m_\mu \rightarrow gA^m_\mu$, we can see
the expansion parameter is  $g^2$, which has a dimension of mass. The
dimensionless coupling constant turns out to be $1/|\kappa|$, which
should be small for the perturbative expansion to work.
With the  usual covariant gauge fixing term,
$$
{\cal L}_{gf} = -{1\over 2\xi } (\partial^\mu A^m_\mu)^2, \eqn\Lgf
$$
we get the Fadeev-Popov ghost Lagrangian,
$$
{\cal L}_{FP}
= \partial^\mu \bar{\eta}^m \partial_\mu \eta^m
+ f^{lmn}  \partial^\mu \bar{\eta}^l  A^m_\mu \eta^n. \eqn\Lgh
$$

Combining the Lagrangian $\OnShL$ and the gauge fixing terms $\Lgf$ and
$\Lgh$, we obtain the quadratic terms
$$
\eqalign{ {\cal L}_0
&\ = {1\over 2g^2} A^{m\mu}\left\{
(\partial^2 \eta_{\mu\nu} -  \partial_\mu \partial_\nu)
- m \epsilon_{\mu\nu\rho} \partial^\rho
+ {1\over \xi}\partial_\mu \partial_\nu \right\} A^{m\nu} \cr
&\ + {1\over 2g^2}C_a \left\{ -\partial^2 - m^2 \right\}C_a
+ {1\over 2g^2}\bar{\lambda}_a
\left\{i\psl - m \right\} \lambda_a
+ {1\over 2g^2}\bar{\chi}
\left\{i\psl + m \right\} \chi \cr
&\ + \bar{\eta}^m (-\partial^2 )\eta^m, \cr}
\eqn\Lquad
$$
where $m \equiv \kappa g^2$.
{}From the quadratic terms $\Lquad$, it is straightforward to get the
propagators for the gauge, scalar, and fermion fields.
Here we  show only the propagator for the gauge field,
$$
\eqalign{(\Delta_0)_{\mu\nu}(p)
&\ = {-ig^2 \left(p^2\eta_{\mu\nu} - p_\mu p_\nu \right)
+ mg^2\epsilon_{\mu\nu\rho} p^\rho \over p^2(p^2- m^2)}
- {i \xi p_\mu p_\nu \over p^4}. \cr}
\eqno\eq
$$
To avoid the infrared divergence, we use the Landau gauge $\xi=0$.

Since the theory is finite, we can calculate perturbatively the
effective action in terms of the bare fields and parameters without
introducing the counter terms. Because of the Lorentz and gauge
invariance, the gluon self-energy takes the form:
$$
\pi_{\mu\nu}(p) = (p^2\eta_{\mu\nu} - p_\mu p_\nu) \Pi_e(p^2)
-i \epsilon_{\mu\nu\rho}p^\rho  \Pi_o(p^2) + p_\mu p_\nu \Pi_3(p^2). \eqno\eq
$$
The kinetic term in the effective action for the gauge boson is then
$$   i(\Delta_{\mu\nu}(p))^{-1} =  i(\Delta_{0\mu\nu}(p))^{-1} +
\pi_{\mu\nu}(p).  \eqno\eq $$
With the correction $i p^2\tilde{\Pi}(p^2)$ to the ghost propagator,
we get the corrected ghost kinetic term,
$$  i(\tilde{\Delta}(p))^{-1} =   i(\tilde{\Delta}_0(p))^{-1} +p^2
\tilde{\Pi}(p).
\eqno\eq $$

The part of the effective action which is similar to the classical
Lagrangian can be written in terms of the renormalized fields and
parameters with the standard normalization. This leads the relation
between the renormalized fields and parameters and the bare fields and
parameters.  For example,
$$\eqalign{&\ A_{\mu}^m = \sqrt{Z_3}\, A_{{\rm ren}\; \mu}^m,  \cr
&\  \eta^m = \sqrt{\tilde{Z}}\, \eta_{\rm ren}^m. \cr}
\eqno\eq $$
{}From Eq.(11)  we see  the ghost field renormalization factor is
$$ \tilde{Z} = 1 - \tilde{\Pi}(0).
\eqno\eq $$
With our normalization of the gauge field kinetic term, the
interaction term between the ghost fields and the vector boson should
be again unity after renormalization by the Ward indentity, which means that
$$ Z_3 = \tilde{Z}^{-2} .\eqno\eq $$
With the definition $Z_k \equiv  1 - \Pi_0(0)/\kappa $, the renormalized
Chern-Simons coefficient is
$$ \eqalign{ \kappa_{\rm ren} &\ = \kappa Z_k Z_3 = \kappa Z_k
\tilde{Z}^{-2} \cr
&\ = \kappa ( 1-{1\over \kappa} \Pi_o (0) +2 \tilde{\Pi} (0) ). \cr}
\eqno\eq  $$

The renormalized coupling $g^2_{\rm ren} = g^2 (Z_g Z_3)^{-1} $ with $Z_g = 1
+ g^2 \Pi_e(0)$ is not much interesting.  The renormalized mass for
the vector bosons can be obtained from  Eq.(9).  We calculate the pole of
the propagator at $p^2=m^2$ and so
$$ m_{\rm ren}   = \bigl[ 1+
g^2\Pi_e(m^2) - {1\over \kappa}\Pi_o(m^2) \bigr]m .
\eqno\eq $$
We can also calculate the similar mass corrections to the matter
fields. As the supersymmetry is preserved, the renormalized mass for
the matter fields will turn out to be  identical to that of the
vector bosons.

We first calculate the correction to the ghost propagator
$$ \tilde{\Pi}(p^2) = {ic_v m \over \kappa p^2} \int {d^3 k \over
(2\pi)^3} {\bigl[ k^2p^2-(k\cdot p)^2 \bigr] \over k^2
(k^2-m^2)(k-p)^2 } .
\eqno\eq $$
There are seven Feynman graphs contributing to the gluon self-energy,
but only three of them yield non-vanishing corrections to $\Pi_o$: one
gluon loop and two fermion loops.  Because of the
supersymmetry, the result is free of ultraviolet divergence and there
is no need of regularization.    After some algebra we have the
bosonic and fermionic contributions,
$$
\eqalign{
& \Pi_o^B(p^2) =  {ic_v m\over p^2}
\int {d^3k\over (2\pi)^3}
 {\bigl[ k^2 p^2 -(k\cdot p)^2 \bigr]
\bigl[ 5k^2-5k\cdot p + 4p^2 -2m^2 \bigr] \over
k^2(k^2-m^2)(k-p)^2\bigl[ (k-p)^2-m^2 \bigr]} , \cr
& \Pi_o^F(p^2) =  {ic_v m\over p^2}
\int {d^3k\over (2\pi)^3} { -2p^2 \over (k^2-m^2)
\bigl[ (k-p)^2 - m^2 \bigr] }, \cr}
\eqno\eq $$
which we can evaluate explicitly. Since $\Pi(p^2)$ and $\tilde{\Pi}_o(p^2)$
vanishes at
the large momentum limit, the parameter $\kappa$  does not get renormalized in
short
distance.   The one-third of the $\lambda_a$
contribution is cancelled by the $\chi$ contribution. The bosonic
contribution to the odd part comes only from the gauge field.
We can see that both $\Pi^B_o$ and $ \Pi^F_o$ are finite.

Taking the zero momentum limit, we get for the $N=3$  case that
$$
\eqalign{ &\tilde{\Pi}(0) = {- c_v  \over 6\pi |\kappa|}, \cr
& \Pi^B_o(0) = {-7c_v \kappa \over 12\pi |\kappa|}, \cr
& \Pi^F_o(0) = {c_v \kappa \over 4\pi |\kappa|}. \cr}
\eqno\eq
$$
The above corrections hold equally well for the $N=2$ case.
But for the
$N=1$ case, the fermionic contribution $\Pi_o^F(0)$ should be one half
of the above result.  From Eq.~(15), we see immediately see that
$$
\eqalign{ &\ \bigl. \kappa_{\rm ren}\bigr|_{N=2,3} = \kappa,
 \cr
&\ \bigl. \kappa_{\rm ren}\bigr|_{N=1} = \kappa + {c_v \kappa  \over
8\pi |\kappa|} ,\cr}
\eqno\eq
$$
to one loop order.  The shift in $k= 4\pi \kappa$ for the $N=1$ case
is $c_v/2$, which is a half integer if $c_v$ is odd. In this case, the
theory would be quantum mechanically anomalous.

For the mass correction, we calculate the $\Pi_e(p^2)$ for the $N=3$
case, which is the sum  of the bosonic contribution $\Pi_e^B(p^2)$ and
the fermionic contribution $\Pi_e^F(p^2)$:
$$
\eqalign{
\Pi_e^B(p^2) = {ic_v\over 2p^4}
\int {d^3k\over (2\pi)^3}
& \biggl\{ {-6\bigl[ k^2 p^2 -(k\cdot p)^2 \bigr]
\over (k^2-m^2)\bigl[ (k-p)^2-m^2 \bigr]} + {6p^2 \over (k^2 - m^2)} \cr
& + {\bigl[ k^2 p^2 -(k\cdot p)^2 \bigr] Q
\over k^2(k^2-m^2)(k-p)^2\bigl[ (k-p)^2-m^2 \bigr]} \cr
& + {(8/3)p^2 \over (k^2 - m^2)} \biggr\}, \cr
\Pi_e^F(p^2) =  {ic_v\over 2p^4}
\int {d^3k\over (2\pi)^3}
& \biggl\{ {-8(k\cdot p)^2 + 8(k\cdot p)p^2 + 8m^2p^2
\over (k^2-m^2)\bigl[ (k-p)^2-m^2 \bigr]} \biggr\}, \cr}\eqno\eq
$$
where  $Q \equiv -3k^2(k-p)^2 - 2k^2p^2 - 4(k-p)^2p^2 - 2(k\cdot p)^2
+ m^2k^2 -m^2(k\cdot p).$  The first line in $\Pi_e^B(p^2)$ comes from the
scalar loop diagrams, the second line comes from the combination of the gluon
and ghost loop diagrams. and the third line comes from the gluon tadpole
diagram.  Note that in $\Pi_e^F(p^2)$, the contributions from $\lambda$
and $\chi$ loops are of the same sign, unlike the case in $\Pi_o^F(p^2).$
The corresponding $N=2$ results can be obtained by multiplying a factor of
$1/3$ to the scalar diagrams and a factor of $1/2$ to the fermion diagrams.
The results for $N=1$ is obtained by dropping the scalar contribution
and multiplying a factor $1/4$ to the fermion contribution. One can
see easily that the linear divergences in $\Pi^B_e$ and $\Pi^F_e$ cancel
each other for each $N$.

We can calculate explicitly $\Pi_e(p^2) $ and $\Pi_o(p^2)$ and get the
results for the $N=3$ case,
$$ \eqalign{  g^2\Pi_e(m^2) &\ = {g^2 c_v \over 64\pi |m|} ( 5\ln 3 - 4 +
i \pi ),  \cr
\Pi_o(m^2)/\kappa &\ = -{g^2 c_v \over 64 \pi |m|} ( 11 \ln 3 + 4 -
i\pi).   \cr}
\eqno\eq $$
Note that the pole of the propagator is $p^2 = m^2 - i\epsilon$.
Similar results can be obtained for $N=1,2$ cases.
 From Eq.(16), we get the mass corrections,
$$
\eqalign{
& \bigl. \Delta m\bigr|_{N=3} = {c_v m\over 4\pi |\kappa|} , \cr
& \bigl. \Delta m\bigr|_{N=2} =  {3c_v m\over 8\pi |\kappa|} , \cr
& \bigl. \Delta m\bigr|_{N=1} =  {c_v m\over 2\pi |\kappa|} . \cr}
\eqno\eq $$
As a check to the above results, we  also calculated the mass
corrections  to all other particles in the supermultiplets  and got
the identical results. We expect the supersymmetry is preserved
quantum mechanically  and see that
the  supersymmetry is preserved as far as the one-loop mass
correction is concerned.

\vskip 0.2in

We thus have obtained the shift of the Chern-Simons coefficient for
the supersymmetric Yang-Mills Chern-Simons theories. For the  $N=2,3$
cases, there is no correction and for the $N=1$ case, the shift is $k
\rightarrow k + c_v/2$. If the correction in the $N=1$ case is a half
integer, the theory would be anomalous. The result depends crucially on the
sign
of the various terms. As a check, we also calculated  the mass correction
 to charged particles and have shown that the supersymmetry is preserved.

There are many interesting implications arising from our results.
It would be interesting to understand the possibly  anomalous $N=1$
supersymmetric theory  in terms of the effective action induced by
the integrating over a massive Majorana fermion. This seems quite
analogous to the global $Z_2$ anomaly discovered by Witten in the four
dimensional gauge theory[8].

In calculating the knot invariant, Witten observed that the shift
$\kappa \rightarrow \kappa + c_v$ occurs  naturally in the pure
Chern-Simons theory[9]. It would be also interesting to find the similar
arguments for the supersymmetric knot invariants which is consistent
with our results.

Finally, our work is also motivated partially by the recent controversy about
the quantum corrections to the Chern-Simons coefficient when the gauge symmetry
is
spontaneously broken by the Higgs field[10].
The calculations done for  both abelian and
nonabelian gauge groups suggest that in the broken phase the quantum
correction depends on whether the coefficient is for the unbroken
symmetry generators or the broken symmetry generators. It seems that
the correction for the unbroken generators is an integer[11] and that for the
broken generators is a complicated function of the coupling constants.
It has been argued that the corrections to the broken generators can be
viewed as parts of a gauge invariant effective action which imitates the
Chern-Simons term in the broken phase, which has been shown explicitly
in abelian case[12].

However, there are good reasons for the correction to the Chern-Simons
coefficient to be quantized even in broken phase[13].  Especially we have
shown that this is indeed the case in a special class of theories, the
abelian self-dual Chern-Simons-Higgs models[13].  We expect also that for
the nonabelian self-dual Chern-Simons-Higgs systems the correction is
quantized.  In calculating the quantum corrections in these theories, one
has to understand the role of the Yang-Mills term and the regularization
dependency.  This work answers this question.

\vskip 0.5in

\centerline{\bf Acknowledgement }

The work by K.L.  is supported in part by U.S.  DOE, the NSF
Presidential Young Investigator program and the Alfred P.  Sloan
Foundation.  The work by T.L.  is supported by Korea Science and
Engineering Foundation (through the Center for Theoretical Physics,
SNU) and the Basic Science Research Institute Program, Ministry of
Education, Korea (No.  BSRI-94-2401).

\vfill \endpage

\refout

\end

\REF\rKhleb{S.Y.  Khlebnikov, JETP letters {\bf 51}, 81, (1990); V.P.
Spiridonov.  \pl{247}{337}{90}.  } 

\REF\rKhlebn{ S.Y.  Khlebnikov and M.E.  Shaposhnikov,
\pl{254}{148}{91}; A.  Khare, R.B.  MacKenzie, P.K.  Panigrahi and
M.B.  Parajape, ``Anomalies via spontaneous symmetry breaking in 2+1
dimensions,'' Univ.  de Montr\'eal Preprint, UdeM-LPS-TH-93-150,
hep-th/9306027}

\REF\rHong{ J.  Hong, Y.  Kim and P.Y.  Pac, \prl{64}{2230}{90}; R.
Jackiw and E.J.  Weinberg, \ibid{64}{2234}{90}; R.  Jackiw, K.  Lee
and E.J.  Weinberg, \pr{42}{3488}{90}; Y.  Kim and K.  Lee,
\ibid{49}{2041}{94}.}

\REF\rClee{C.  Lee, K.  Lee, and E.J.  Weinberg, \pl{243}{105}{90};
E.A.  Ivanov,\ibid{268}{203}{91}; S.J.  Gates and H.  Nishino,
\ibid{281}{72}{92}.}

\REF\rKao{H.-C.  Kao and K.  Lee, \pr{46}{4691}{92}; H.-C.
Kao,``Self-Dual Yang-Mills Chern-Simons Higgs Systems with an N=3
Extended Supersymmetry,'' Columbia preprint, CU-TP-595 (1993), to
appear in Phys.  Rev.  D.}

\REF\rKlee{ K.  Lee, \prl{66}{553}{91}; K.  Lee, \pl{255}{381}{91};
G.V.  Dunne, \ibid{324}{359}{94}; H.-C.  Kao and K.  Lee, ``Self-Dual
SU(3) Chern-Simons Higgs Systems,'' Columbia preprint, CU-TP-635
(1994); G.  Dunne, ``Vacuum Mass Spectra for SU(N) Self-Dual
Chern-Simons-Higgs Systems,'' Univ.  Connecticut preprint UCONN-94-4
(1994).}

\REF\rIpeko{ Y.  {\.  I}peko{\v g}lu, M.  Leblanc and M.T.  Thomaz,
Ann.  Phys. (N.Y.), {\bf 214}, 160 (1992).}